\documentclass[11pt,preprint]{aastex}
\newcommand \versnum {8}
\usepackage{color}
\usepackage{natbib}
\usepackage{amsmath}
\tightenlines
\slugcomment{}

\newcommand \Angstrom   {\,{\rm \AA}}

\newcommand \beq        {\begin{equation}}
\newcommand \beqa	{\begin{eqnarray}}

\newcommand \cm         {\,{\rm cm}}

\newcommand \Debye      {\,{\rm D}}

\newcommand \eeq	{\end{equation}}
\newcommand \eeqa	{\end{eqnarray}}

\newcommand \gm         {\,{\rm g}}
\newcommand \GHz        {\,{\rm GHz}}
\newcommand \gtsim	{\gtrsim}		 

\newcommand \kB         {k_{\rm B}}

\newcommand \K  	{\,{\rm K}}

\newcommand \ltsim	{\lesssim}		 
\newcommand \lya        {Lyman-$\alpha$}

\newcommand \mH         {m_{\rm H}}
\newcommand \MHz        {\,{\rm MHz}}

\newcommand \nH         {n_{\rm H}}

\newcommand \NH         {N_{\rm H}}

\newcommand \Trot       {T_{\rm rot}}

\newcommand \xtimes     {{\!\,\times\!\,}}
\newcommand \Volt       {{\rm Volt}}

\newcommand{\newtext}[1]{{#1}}

\countdef\decade=200
\advance\decade by \year
\countdef\hours=201
\advance\hours by \time
\divide\hours by 60
\countdef\mins=202
\advance\mins by \hours
\multiply\mins by 60
\multiply\hours by 100
\countdef\miltime=203
\advance\miltime by \hours
\advance\miltime by \time
\advance\miltime by -\mins
\renewcommand\today{\number\decade.\number\month.\number\day.\number\miltime}
\begin{document}

\title{%
        \vspace*{-3.0em}
        {\normalsize\rm {\it The Astrophysical Journal Letters}, {\bf 858}:L10 
        (2018 May 10)}\\ 
        \vspace*{1.0em}
        Absorption by Spinning Dust: a Contaminant for
 High-Redshift 21\,cm Observations
	}

\author{B. T. Draine\affil{Princeton University Observatory,
Peyton Hall, Princeton, NJ 08544-1001, USA; draine@astro.princeton.edu}
Jordi Miralda-Escud\'e\affil{Institut de Ci\`encies del Cosmos, Universitat
de Barcelona (IEEC-UB), 08028 Barcelona, Catalonia, Spain;
miralda@icc.ub.edu}
}

\begin{abstract}
Spinning dust grains in front of the bright
Galactic synchrotron background can produce a weak absorption signal
that could affect measurements of high--redshift 21\,cm absorption.
At frequencies near 80 MHz where the Experiment to Detect
the Global EoR Signature (EDGES) experiment
has reported 21\,cm absorption at $z\approx17$, absorption could be
produced by interstellar nanoparticles with radii $a\approx50\Angstrom$ 
in the cold interstellar medium (ISM),
\newtext{with rotational} temperature $T\approx 50$ K.
Atmospheric aerosols could 
contribute additional absorption.
The strength of the absorption depends on the abundance of such
grains and on their dipole moments, which are uncertain. The
breadth of the absorption spectrum of spinning dust limits its possible
impact on measurement of a relatively narrow 21\,cm
absorption feature.
\end{abstract}

\section{Introduction}

\citet{Bowman+Rogers+Monsalve+etal_2018} report a small decrease in
the sky brightness in the 50--100\,MHz region that has been interpreted
as arising from 21\,cm absorption of the cosmic microwave background (CMB) 
at redshift $z\sim17$.
The reported strength of the absorption signal, however, exceeds by a
factor $\sim 3$ the largest model predictions based on an atomic intergalactic
medium (IGM) that cools adiabatically after electron Thompson scattering can
no longer keep the kinetic temperature coupled to the CMB temperature.
This discrepancy has led to new physics being proposed, such as elastic
scattering of dark matter by baryons or electrons that could lower the
kinetic temperature of the baryons and possibly lead to a lowering of
the spin temperature \citep{Barkana_2018}. This hypothesis, 
not motivated by theory or by independent
observations, faces an additional difficulty: if the \lya\ photons
emitted by the first stars that are necessary to couple the spin and
kinetic temperature of the hydrogen gas through the Wouthuysen--Field
effect are accompanied by as little as $\sim 1\%$ of their energy in
X-ray emission or cosmic ray production, the resulting heating of the
intergalactic gas would reduce the maximum strength of
the absorption signal by a factor $\sim 10$ 
\citep{Chen+Miralda-Escude_2004,Hirata_2006}.
The first population of stars would therefore need to produce an
anomalously low emission of X-rays and cosmic rays due to supernova
remnants and X-ray binaries, compared to local starbursts 
\citep{Oh_2001,Pacucci+Mesinger+Mineo+Ferrara_2014}.

Observations of the CMB in this frequency range must contend with
bright Galactic synchrotron emission with brightness temperature
ranging over $1500 \lesssim T_{\rm B} \lesssim 4000\K$ at 78 MHz, so any
frequency-dependent absorption (Galactic or telluric) interposed between
us and the synchrotron emission will 
need to be recognized and, if significant, corrected for.
For standard models, the expected signal 
due to 21\,cm absorption at redshift $z\approx17$
is at most $\Delta T_{\rm B} \approx -0.2\K$ 
\citep{Zaldarriaga+Furlanetto+Hernquist_2004}.
Against a $T_{\rm B}\sim 2000\K$ background, an absorption optical depth
$\tau\sim 10^{-4}$ would produce a signal of this magnitude.

Interstellar dust models \citep[e.g.,][]{Draine+Li_2007}
already include substantial populations of
nanoparticles in order to explain the observed infrared emission, 
including 
strong emission features at 3.3, 6.2, 7.7, 8.6, 11.3, and 12.7$\micron$
attributed to polycyclic aromatic hydrocarbon (PAH) particles
\citep{Tielens_2008},
as well as the ``anomalous microwave emission'' 
\citep{Dickinson+Ali-Haimoud+Barr+etal_2018} 
that has been
interpreted as rotational emission from $a\approx5\Angstrom$
nanoparticles spinning at $\sim15-60\GHz$
\citep{Draine+Lazarian_1998a}.
Here we explore the possibility that spinning interstellar dust grains
with radii $a\approx 50\Angstrom$ could produce
absorption in the $20-200\MHz$ frequency range.
Rotational 
absorption by telluric aerosols could also affect ground-based observations.
We find that 
absorption by interstellar dust or atmospheric aerosols
is too broad to reproduce the signal reported
by \citet{Bowman+Rogers+Monsalve+etal_2018}.  
However, this process (absorption of
the bright Galactic synchrotron radiation by spinning dust)
should be considered as a possible 
astrophysical ``foreground'' that could contaminate CMB observations
in this frequency range.

\section{\label{sec:rotational absorption}
         Rotational Absorption by Spinning Dust}

Consider grains of \newtext{mass $M$ 
and number density $n_{\rm gr}$,
in ``Brownian rotation'' at temperature $\Trot$, with an electric
dipole moment $\mu_\perp$ perpendicular to the rotation axis.
The volume-equivalent radius is
$a=(3M/4\pi\rho)^{1/3}$, where $\rho$ is the density.   
The moment of inertia is $I=\alpha (2/5) Ma^2$, where
$\alpha=1$ for a solid sphere.}
The total power emitted by one grain rotating at
frequency $\nu$ is
\beq
P =  \frac{2}{3}\frac{\mu_\perp^2}{c^3} (2\pi\nu)^4 ~.
\eeq
The density of grains rotating at frequency $\nu$ within a range $d\nu$
is
\beq \label{eq:thermal sphere}
f(\nu)\, d\nu = n_{\rm gr} \,
\frac{4\,\alpha^{3/2} }{\sqrt{\pi}\, \nu_T^3}\,
e^{-\alpha(\nu/\nu_T)^2}\, \nu^2 d\nu ~.
\eeq
The frequency $\nu_T$ is related to $T_{\rm rot}$ by
\beqa \label{eq:nu_T}
\nu_T &\equiv& \left(\frac{15 \kB\Trot}{16\pi^3\rho a^5}\right)^{1/2}
= 58 
\left(\frac{\Trot}{50\K}\right)^{1/2}
\left(\frac{2\gm\cm^{-3}}{\rho}\right)^{1/2}
\left(\frac{50\Angstrom}{a}\right)^{5/2} \MHz
~.
\eeqa
The power radiated in frequency interval $d\nu$ per unit volume and
sterradian is then
\beq \label{eq:j_nu}
j_\nu d\nu = \frac{n_{\rm gr}}{4\pi}\,
 \frac{2}{3}\frac{\mu_\perp^2}{c^3} \,
(2\pi\nu)^4
\,
\frac{4\alpha^{3/2}}{\sqrt{\pi} \nu_T^3}
\,
e^{-\alpha(\nu/\nu_T)^2}
\,
\nu^2 d\nu  ~.
\eeq
The effective absorption cross section (i.e., true absorption minus
stimulated emission) is obtained 
from Kirchoff's law (in the Rayleigh-Jeans limit
$B_\nu(T_{\rm rot}) = 2kT_{\rm rot} \nu^2/c^2)$,
\beqa
n_{\rm gr}C_{\rm abs}(\nu) &=& \frac{j_\nu}{B_\nu(\Trot)}
\\ \label{eq:ngrCabs}
&=&
n_{\rm gr}
\frac{2^{10}\pi^7}{45\sqrt{15}}
\frac{\mu_\perp^2}{c}
\frac{\rho^{3/2} a^{15/2}}{(\kB\Trot)^{5/2}}
\alpha^{3/2}\,
\nu^4
e^{-\alpha(\nu/\nu_T)^2}
~.
\eeqa 
This cross section $C_{\rm abs}(\nu)$ peaks at a frequency
\beq
\nu_{\rm peak} = \sqrt{\frac{2}{\alpha}}\,\nu_T = \frac{82}{\sqrt{\alpha}}
\left(\frac{\Trot}{50\K}\right)^{1/2}
\left(\frac{2\gm\cm^{-3}}{\rho}\right)^{1/2}
\left(\frac{50\Angstrom}{a}\right)^{5/2} \MHz
~.
\eeq

\newtext{
The $\nu^4e^{-\alpha(\nu/\nu_T)^2}$ dependence 
of the opacity in Eq.\ (\ref{eq:ngrCabs})
is for the thermal distribution 
of Eq.\ (\ref{eq:thermal sphere}),
exact for spheres.
The cutoff frequency is $\nu_T \propto \sqrt{\Trot/I}$, where
$I$ is the moment of inertia.
For more realistic grain shapes, the moment of inertia tensor has
eigenvalues $I_1 > I_2 > I_3$, and the distribution of
rotation frequencies is complicated.
Even a single grain with fixed angular momentum $J$ and fixed
rotational kinetic
energy $E_{\rm rot}$ has multiple frequencies characterizing its tumbling
\citep[see, e.g.,][]{Weingartner+Draine_2003,Hoang+Lazarian+Draine_2011}.
The rotational dynamics are further complicated if the grain's vibrational
temperature $T_{\rm vib} < \Trot$, as is generally the case for interstellar
grains: the particle then tends
to align its principal axis with largest
moment of inertia with the angular momentum vector.

Define $\alpha_j\equiv I_j/(\frac{2}{5}Ma^2)$.
For example, an ellipsoid with axial ratios $1$:$\sqrt{2}$:$2$ has
$(\alpha_1,\alpha_2,\alpha_3)=(1.5,1.25,0.75)$.
An exact description of rotation for triaxial shapes
is beyond the scope of the present
work, but a simple approximation is to average over
thermal distributions for spheres for the three $\alpha_j$:
\beqa \label{eq:triaxial}
f(\nu)d\nu
&\approx& n_{\rm gr} \frac{4}{\sqrt{\pi}} \nu^2d\nu \times
\frac{1}{3}
\sum_{j=1}^3 \frac{\alpha_j^{3/2}}{\nu_T^{3}}e^{-\alpha_j (\nu/\nu_T)^2}
\\
n_{\rm gr}C_{\rm abs}
&\approx& 
n_{\rm gr} \frac{2^{10}\pi^7}{45\sqrt{15}}
\frac{\mu_\perp^2}{c}
\frac{\rho^{3/2}a^{15/2}}{(k\Trot)^{5/2}} \nu^4 
\times \frac{1}{3}
\sum_{j=1}^3 \alpha_j^{3/2}e^{-\alpha_j (\nu/\nu_T)^2}
~.
\eeqa
The actual distribution for a triaxial body in thermal equilibrium
is expected to be intermediate between 
Eq.\ (\ref{eq:triaxial}) and the simple thermal distribution 
(\ref{eq:thermal sphere}) for a single
$\bar{\alpha}=(\alpha_1+\alpha_2+\alpha_3)/3$.
We will compare these below.
}

\section{\label{sec:size dist}
         Nanoparticle Abundances}

The composition and size distribution of interstellar grains remain uncertain.
Models to reproduce the optical and UV
extinction require a size distribution with most of the grain mass
concentrated in the $0.05$--$0.5\micron$ size range.
However, a substantial additional population of smaller grains is required
to account for the ultraviolet extinction.
A significant fraction of the smallest particles, containing as few
as $\sim$50 atoms, must be composed of polycyclic aromatic hydrocarbons
(PAHs) in order to account for the observed strong infrared emission 
bands at 3.3, 6.2, 7.7, 8.6, 11.3, and 12.7$\micron$. 
In the model of \citet{Draine+Li_2007}, the PAH nanoparticles are concentrated
in two populations -- one with the mass distribution peaking
near $6\Angstrom$, containing $\sim$$15\%$ of the interstellar carbon,
and one peaking near $50\Angstrom$, containing $\sim5\%$ of the interstellar
carbon.
The latter population was invoked to account for observed 24$\micron$
continuum emission from the ISM of spiral galaxies.

If composed primarily of carbon 
(perhaps PAHs), 
these particles would have a mean density $\rho\approx 2\gm\cm^{-3}$,
and a number of atoms $N \approx 5\times10^4 (a/50\Angstrom)^3$.
Nanoparticles with other compositions, such as silicate, may also be
numerous without violating observational constraints
\citep{Hensley+Draine_2017b}.
A substantial fraction of the iron in the ISM could
be in the form of metallic Fe nanoparticles
\citep{Hensley+Draine_2017a}, iron oxides, or other material.

The interstellar grain size distribution is the result of a balance
between grain growth (by accretion and coagulation), and
grain destruction (by vaporization and fragmentation in grain-grain collisions,
sputtering in hot gas, and photodestruction of the smallest grains).
It is at least conceivable that the larger $\gtsim0.1\micron$
interstellar grains might
consist of domains containing $\sim 10^4-10^5$ atoms, with
shattering producing a fragmentation spectrum peaking in this size range.
As noted above, models aiming to reproduce the infrared emission 
\citep{Draine+Li_2007} specifically
included a component peaking at this size. A peak in the
size distribution would result in a peak in the spinning dust
absorption spectrum.

\section{\label{sec:dipole moment}
         Nanoparticle Dipole Moments}

If the rotation axis is uncorrelated with the
direction of the electric dipole moment,
then
\beq \label{eq:2/3}
\langle \mu_\perp^2 \rangle = \frac{2}{3}\mu^2
~,
\eeq
where $\mu$ is the total electric dipole moment of the grain. 
We will be primarily interested in
nanoparticles with radii $a\approx 50\Angstrom$.
Several different effects may contribute to $\mu$.

{\it Amorphous materials:}
an amorphous material can be thought of as a random aggregation
of structural units with no long-range order.
A nanoparticle of amorphous composition might then have an electric
dipole moment $\mu \approx \mu_{\rm unit}\times \sqrt{N/N_{\rm unit}}$ where
$N_{\rm unit}$ is the number of atoms in one structural unit, and
$\mu_{\rm unit}$ is the typical dipole moment of a structural unit.
The aliphatic C--H bond has a dipole moment $0.3\Debye$
(D $\equiv$ Debye  $\equiv10^{-18}{\rm\,esu}\cm$), the Si--H bond has a 
dipole moment $1.0\Debye$ \citep{Speight_2005}, and
the Si--O bond has dipole moment $0.95\Debye$ \citep{Freiser+Eagle+Speier_1953}.
For materials with polar structural units,
values of $\mu_{\rm unit}/\sqrt{N_{\rm unit}} \approx$ few Debye are
plausible, thus for $N\approx 5\xtimes10^4$ one might expect
$\mu \approx 200 \times {\rm few}\Debye$.

{\it Inhomogeneous nanoparticles:}
the nanoparticle may be an aggregation of different materials with
different work functions (i.e., different Fermi levels). 
If the materials are metals, then electrons will flow to develop
a ``contact potential'' between the different components.
To illustrate this, suppose that the nanoparticle consisted of
two conducting hemispheres with a small gap $\delta$ separating
the hemispheres.  To produce a potential difference $\Delta V_{\rm contact}$
between the hemispheres, the electric field in the gap would be
$E_{\rm gap}=\Delta V_{\rm contact}/\delta$, with a charge 
$\pm Q$ on each side of the gap, with
$Q=E_{\rm gap}\times (\pi a^2/4\pi)$.
The electric dipole moment of this ``double layer'' is
\beq
\mu = Q\delta = \left(\Delta V_{\rm contact}\right) \frac{a^2}{4}
= 210 \left(\frac{\Delta V_{\rm contact}}{\Volt}\right)
\left(\frac{a}{50\Angstrom}\right)^2
\Debye
~,
\eeq
independent of $\delta$.
Contact potentials 
$\Delta V_{\rm contact} \approx 1 \Volt$ are plausible.
Thus, one might expect $\mu\approx 200\Debye$ from this effect.

{\it Ionized nanoparticles:}
capture of electrons and photoionization
can result in negatively or positively charged nanoparticles.
If photoionization is faster than electron capture for a neutral
particle, then it will generally be found positively charged,
with a net potential 
$U$, and a net charge 
\beq
Z = \frac{Ua}{e} = 3.5 \left(\frac{U}{\Volt}\right)
\left(\frac{a}{50\Angstrom}\right)
~.
\eeq
If the grain is metallic, the surface will be an equipotential.
\citet{Purcell_1975} noted that for moderately irregular isopotential
surfaces, the center of charge deviates from the centroid by perhaps
$\sim2\%$ of the characteristic radius, in which case the dipole moment
would be only
\beq
\mu \approx 0.02 Zea \approx 16 \left(\frac{U}{\Volt}\right)
\left(\frac{a}{50\Angstrom}\right)\Debye
~.
\eeq
On the other hand, if the material is an insulator, the
excess charge may be localized at a small number of points on the surface.
A single charge $e$ separated from the center of mass
by a distance $5\xtimes10^{-7}\cm$ corresponds to $\mu=240\Debye$.
This dipole moment will, however, be partially
compensated by polarization of the grain material, but the resulting
net dipole moment could plausibly be $\sim 100\Debye$ for a grain with
charge $Z = \pm1$.
Grains with two or three charged sites (which might include both positive
and negative charges) could have larger $\mu$.

{\it Ferroelectric inclusions:}
Some materials become spontaneously electrically polarized when cooled
below a critical temperature, a phenomenon referred to as
ferroelectricity \citep[see, e.g.,][]{Xu_1991}.
Ferroelectric materials are used as
piezoelectrics, computer memory, and liquid crystal displays.
Technologically important materials usually contain cosmically scarce
elements such as Ti or Ba (e.g., BaTiO$_3$).
However, HCl, KNO$_3$ and NH$_4$NO$_3$ are also ferroelectrics, so one should
consider the possibility that interstellar nanoparticles may contain
ferroelectric inclusions.  A low--temperature 
ferroelectric phase of H$_2$O ice has
also been reported \citep{Fukazawa+Hoshikawa+Ishii+etal_2006}.
With dipole moments per atom reaching values
of $\sim1\Debye$, if a fraction $f_{\rm fe}$ of a nanoparticle 
consisted of a ferroelectric domain, the net dipole moment could reach
$\mu \sim f_{\rm fe}N\Debye$, or $\sim10^3\Debye$ for $f_{\rm fe}=2\%$ and
$N=5\times10^4$.

{\it Polarized ice mantles:} ices deposited
on grains can also be spontaneously polarized 
\citep{Field+Plekan+Cassidy+etal_2013,Plekan+Rosu-Finsen+Cassidy+etal_2017}
by either the substrate itself or the electric field if the grain is charged.
This polarization can extend through hundreds of monolayers, but the
observed net polarization per dipole is only a few percent of the
intrinsic dipole moment per molecule.  
If one region on
the surface of a nanoparticle had $\sim10^3$ molecules deposited in
polarized form, this domain might contribute
an electric dipole moment of $\sim10^2\Debye$. 

{\it Ferromagnetism:} interstellar grains may contain
ferrimagnetic or ferromagnetic material
giving the nanoparticle a permanent magnetic dipole moment
\citep[see, e.g.,][and references therein]{Draine+Hensley_2013}.
The rotational emissivity
is still given by Eq.\ (\ref{eq:j_nu}), but with $\mu_\perp^2$
replaced by $(\mu_{e,\perp}^2+\mu_{m,\perp}^2)$ where
$\mu_{e,\perp}$ and $\mu_{m,\perp}$ are the electric and magnetic
dipole moments perpendicular to the rotation axis.
If the grain contains $N_{\rm Fe}$ atoms in a single ferromagnetic domain,
then $\mu_{m}=200(N_{\rm Fe}/10^4)\Debye$, assuming 2.2$\mu_B$ per
Fe atom as in metallic Fe (Bohr magneton $\mu_B = 0.0093\Debye$). 

Each of the above effects leads to dipole moments that might be
as large as
$|\mu| \approx$ few hundred Debye.
As they will act more or less independently, 
overall dipole moments $\mu$ as large as $\sim10^3\Debye$ are not implausible.
Given the uncertainties in the relevant physics, dipole moments as large
as $\sim2000\Debye$ for $a=50\Angstrom$ cannot be ruled out at this time.

\section{\label{sec:excitation}
         Rotational Excitation and Damping}

A number of processes contribute to rotational excitation and damping
of interstellar nanoparticles.
\citet[][see Figs. 4 and 5]{Draine+Lazarian_1998b} compared the importance of 
different
processes as a function of grain size. 
For $a\approx50\Angstrom$ and the $\mu$ values estimated above, 
one can show that damping resulting from
electric dipole radiation, and excitation of rotation by energy
absorbed from the synchrotron background, have negligible effects
on the grain rotation.
For nanoparticles with $a \approx 50\Angstrom$
the rotational \newtext{distribution}
is expected to be approximately thermal, with 
\newtext{
$\Trot\approx50\K$.
For grains of this size, the rotational temperature tends to be lower
than the kinetic temperature of the gas 
because of damping by infrared emission
\citep{Hoang+Lazarian+Draine_2011,Draine+Hensley_2016}.
}

\section{\label{sec:discuss}
         Discussion}

With the absorption cross section from Eq.\ (\ref{eq:ngrCabs}),
dust in the ISM would contribute an optical depth
\beqa
 \label{eq:tau}
\tau_{\rm rot} &=& 
\NH \frac{n_{\rm gr}}{\nH}\, C_{\rm abs}(\nu) =
\NH
X_{\rm gr}
\frac{2^8\pi^6}{15\sqrt{15}}
\frac{\mu_\perp^2\, \mH}{c}
\frac{\rho^{1/2}a^{9/2}}{(\kB\Trot)^{5/2}}
\nu^4
\alpha^{3/2}
e^{-\alpha(\nu/\nu_T)^2}
\\
&=& \nonumber
5.96\xtimes10^{-4} 
\left(\frac{\NH}{10^{21}\cm^{-2}}\right)
\left(\frac{X_{\rm gr}}{4\xtimes10^{-4}}\right)
\left(\frac{\mu}{2000\Debye}\right)^2
\left(\frac{a}{50\Angstrom}\right)^{9/2}\times
\\
&&\hspace*{2.0cm} \label{eq:tau result}
\left(\frac{\rho}{2\gm\cm^{-3}}\right)^{1/2}
\left(\frac{50\K}{T_{\rm rot}}\right)^{5/2}
\left(
\frac{\nu}{100\MHz}
\right)^4
\alpha^{3/2}
e^{-\alpha(\nu/\nu_T)^2}
~,
\eeqa
where we have set $\mu_\perp^2=(2/3)\mu^2$,
\beq
X_{\rm gr} \equiv \frac{n_{\rm gr}}{\nH}
\frac{4\pi\rho a^3}{3 \mH}
\eeq
is the mass in the dust component relative to the mass of H in the ISM,
$\mH$ is the mass of the hydrogen atom,
and $\nu_T$ is given by Eq.\ (\ref{eq:nu_T}).
For all grains in the ISM we have $\sum X_{\rm gr}\approx0.010$
\citep{Draine_2011a}.
If $T_{\rm rot}$ is less than the brightness temperature of the
synchrotron background, the dust will appear in absorption.

For $\NH=10^{21}\cm^{-2}$ 
the ``vibrational'' modes of dust grains are estimated to contribute
$\tau\approx1\xtimes10^{-7}(\nu/100\GHz)^{1.7}$ at $\nu<400\GHz$
\citep{Li+Draine_2001b},
extrapolating to $\tau\approx10^{-12}$ at $100\MHz$.
Absorption by aligned spins in ferromagnetic grains \citep{Draine+Hensley_2013}
could increase this to $\sim10^{-9}$ if most of the Fe is in metallic form,
but this is still completely negligible.

\begin{figure}[t]
\begin{center}
\includegraphics[angle=0,width=8.0cm,
                 clip=true,trim=0.5cm 5.0cm 0.5cm 2.5cm]
{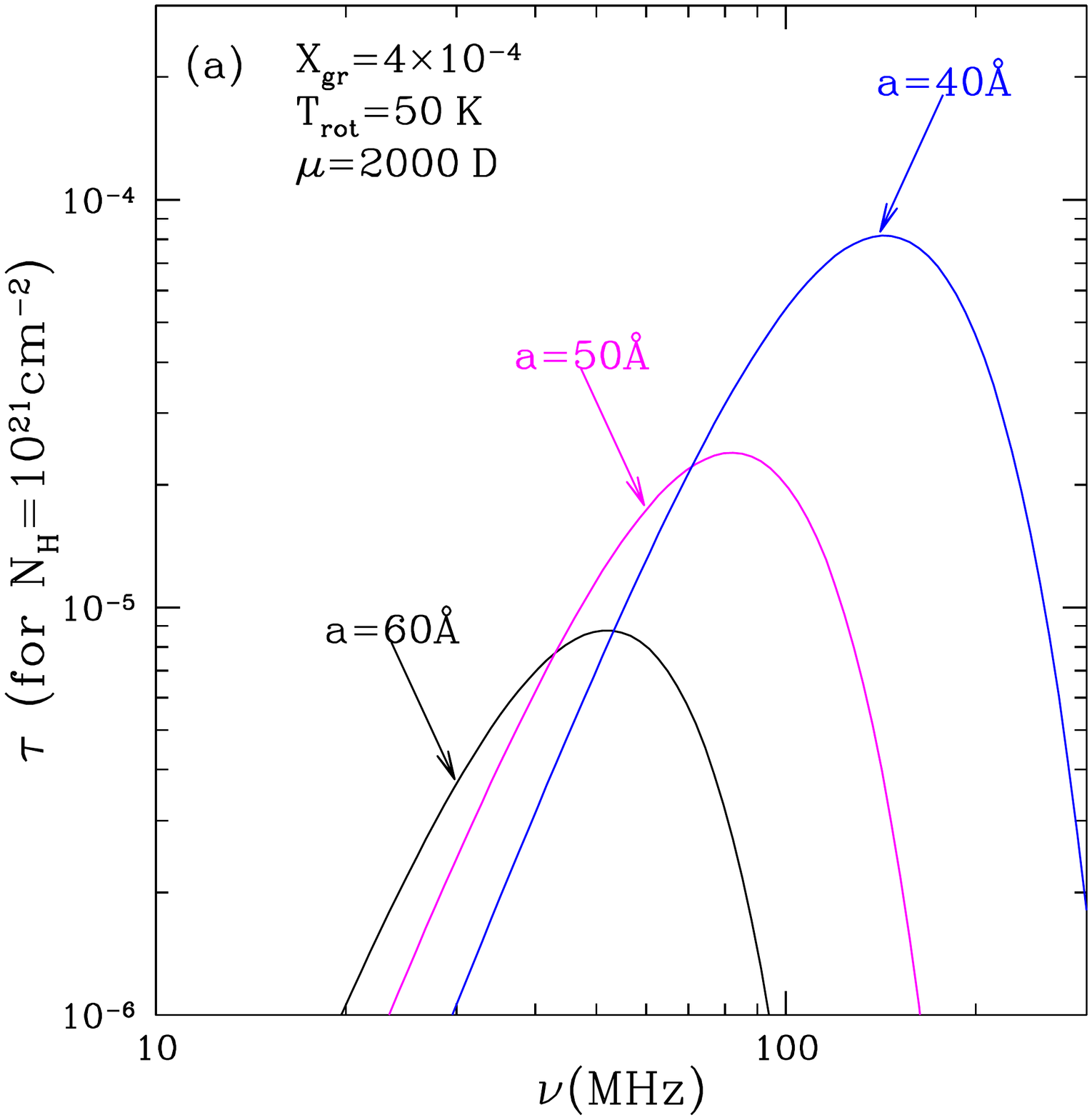}
\includegraphics[angle=0,width=8.0cm,
                 clip=true,trim=0.5cm 5.0cm 0.5cm 2.5cm]
{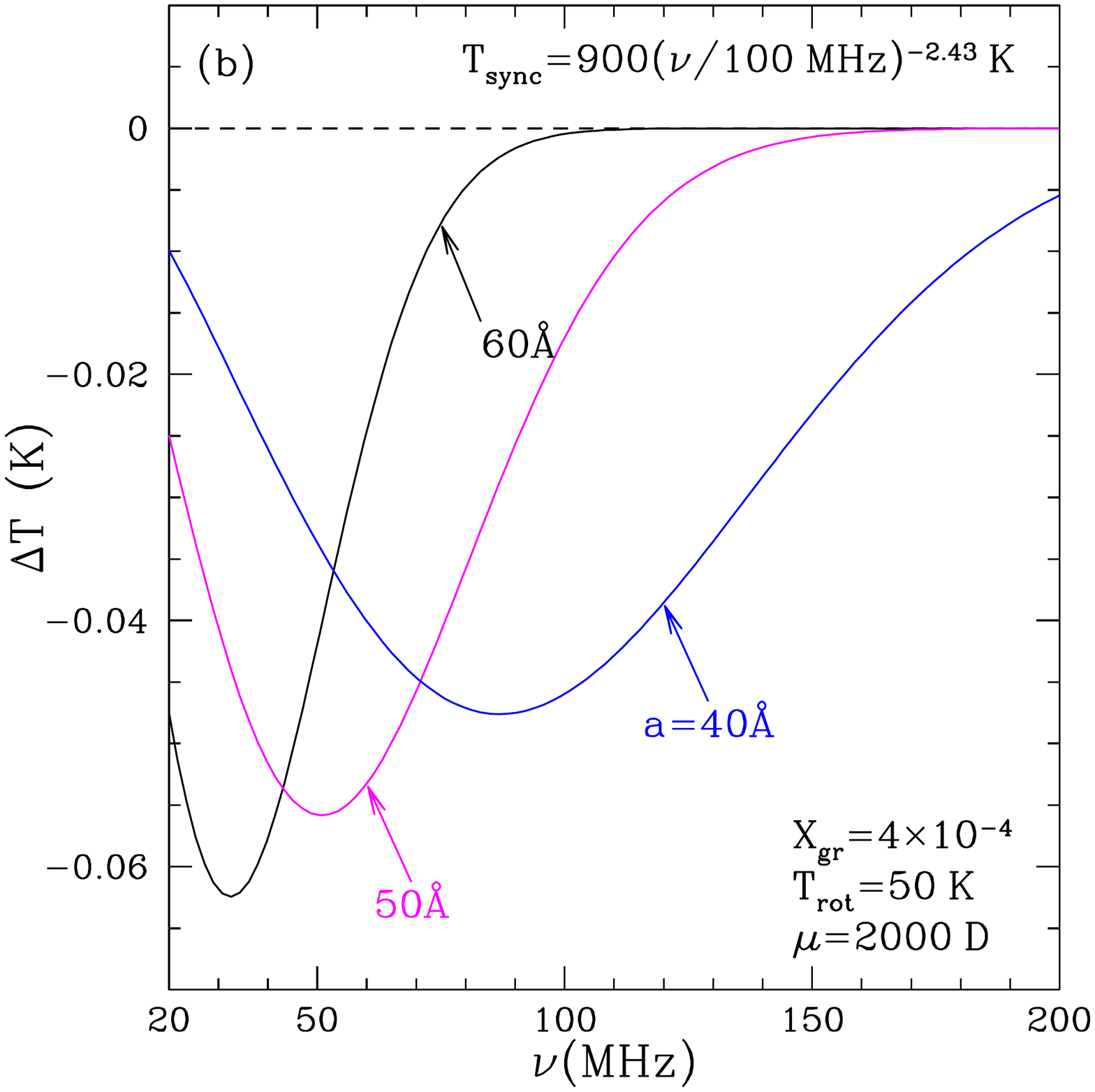}
\caption{\label{fig:tau} \footnotesize
(a) Optical depth produced by nanoparticles in
a column $\NH=10^{21}\cm^{-2}$ of CNM.
Each curve is for nanoparticles of the indicated size with
total mass relative to H
$X_{\rm gr}=4\xtimes10^{-4}$ and $\mu=2000\Debye$.
(b) Absorption signal for assumed synchrotron background
of Eq.\ (\ref{eq:synch}).}
\end{center}
\end{figure}

\newtext{
From Eq.\ (\ref{eq:tau result}) we see that for $\mu\ltsim 100\Debye$,
spinning dust absorption would contribute negligibly to the signal
reported by \citet{Bowman+Rogers+Monsalve+etal_2018}.
However, if we instead take $\mu=2000\Debye$ 
and $a\approx 50\Angstrom$,
we have 
$\tau_{\rm rot}\approx 2\xtimes10^{-5}$ near 80 MHz (see Fig.\ \ref{fig:tau}a).
Such absorption would produce a signal
comparable to the predicted cosmological signal
for realistic scenarios where the harmonic-mean spin temperature
is intermediate between the CMB temperature and the kinetic temperature
of the \ion{H}{1}.
}

\newtext{
If the grains have a single size, the absorption will be peaked as seen
in Figure \ref{fig:tau}a, with $\Delta\nu_{\rm FWHM}=0.82\nu_{\rm peak}$.
For the more realistic case of a size distribution,
the absorption feature 
will be broadened, with $\Delta\nu_{\rm FWHM}/\nu_{\rm peak}>0.82$.
  
Suppose that a column density $\NH=10^{21}\cm^{-2}$ of diffuse ISM
is situated 
in front of a background with brightness temperature
\beq
T_{\rm back} = 2.73\K + T_{\rm sync}
\eeq
where
\beq \label{eq:synch}
T_{\rm sync}= 900 
\left(\frac{\nu}{100\MHz}\right)^{-2.43}\K
\eeq
approximates the synchrotron brightness temperature
reported by \citet{Bowman+Rogers+Monsalve+etal_2018}.
}
The reduction in brightness temperature is
\beq
\Delta T = \left(\Trot-T_{\rm back}\right)\left(1-e^{-\tau}\right)
\approx \left(\Trot-T_{\rm back}\right)\, \tau
~.
\eeq
Figure \ref{fig:tau}b shows the decrease in sky brightness for three
grain sizes.
For $T_{\rm rot}<100\K$, spinning dust would appear in
absorption for $\nu \ltsim 200\MHz$.

\begin{figure}[ht]
\begin{center}
\includegraphics[angle=0,width=8.0cm,
                 clip=true,trim=0.5cm 5.0cm 0.5cm 2.5cm]
{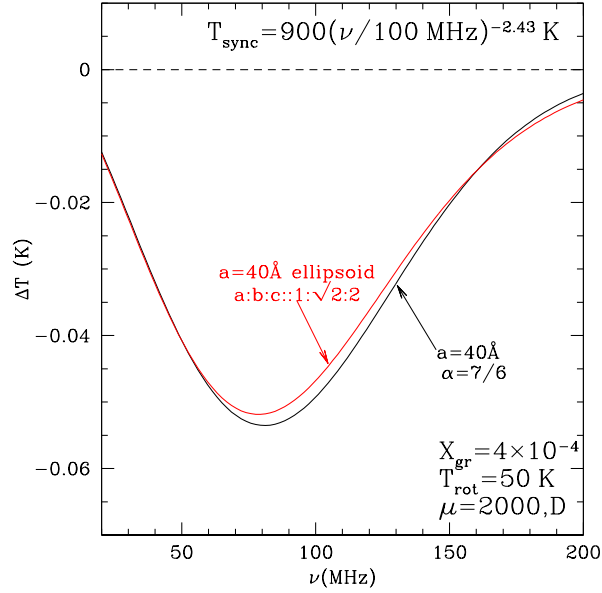}
\caption{\label{fig:tau_ellipsoid} \footnotesize
Absorption by ellipsoidal grains with volume-equivalent radius
$a=40\Angstrom$, estimated from Eq.\ (\ref{eq:triaxial}),
compared to a $a=40\Angstrom$ spheres and $\bar{\alpha}=7/6$.
}
\end{center}
\end{figure}
 

\newtext{
The effect of grain nonsphericity is illustrated in 
Figure \ref{fig:tau_ellipsoid}, comparing absorption 
by ellipsoids with axial ratios
$1$:$\sqrt{2}$:$2$, estimated from
Eq.\ (\ref{eq:triaxial}), to the absorption for spheres of the same mass
and $\bar{\alpha}=7/6$.
Triaxiality does broaden the absorption profile, but the effects
are small compared to the effect of a distribution of grain sizes.}

\newtext{The absorption produced even by single-sized grains is significantly
broader than the absorption profile reported by 
\citet{Bowman+Rogers+Monsalve+etal_2018}.
} 
However, adding a dust absorption component to the background model can
provide more flexibility for fitting the observed shape of the total
spectrum and could reduce the required narrower component that is attributed
to high-redshift 21\,cm absorption.

In addition to interstellar nanoparticles,
atmospheric aerosols could also affect
ground-based observations.
Let $\Sigma$ be the mass surface density in aerosol particles of radius $a$
along the path through the atmosphere.
The optical depth is
\beqa
\tau_{\rm rot} &=& \frac{2\mu^2\Sigma}{3\rho a^3 c}
\left(\frac{15}{\rho a^5\kB T_{\rm rot}}\right)^{1/2}
\left(\frac{\nu}{\nu_T}\right)^4 e^{-(\nu/\nu_T)^2}
\\
&=&1.2\xtimes10^{-4}\left(\frac{\Sigma}{\mu{\rm g}\cm^{-2}}\right)
\left(\frac{\mu}{10^4\Debye}\right)^2
\left(\frac{70\Angstrom}{a}\right)^{11/2}
\left(\frac{\nu}{\nu_T}\right)^4 e^{-(\nu/\nu_T)^2}
~,
\eeqa
where $T_{\rm rot}\approx 250\K$ and $\nu_T$ is given by
Eq.\ (\ref{eq:nu_T}).
For the synchrotron brightness of Eq.\ (\ref{eq:synch}), aerosols
would appear in absorption for $\nu \ltsim 150\MHz$.

Measurements of atmospheric aerosols in Finnish forests 
\citep{Dal_Maso+Kulmala+Riipinen+etal_2005}
show a peak near $a\approx30\Angstrom$, and
a second 
peak near $200\Angstrom$,
but little appears to be known about the abundances \newtext{or compositions}
of very small
aerosols at other locations or altitudes.
\newtext{
Nanoparticles with sizes $a\ltsim 100\Angstrom$ are 
difficult to collect and identify.
Because they are inefficient at scattering optical light,
they do not appear directly as noctilucent
clouds.
In the upper stratosphere, 
such particles are likely to be primarily solids of interplanetary
origin \citep{Flynn_1994}, or condensates from material ablated
from meteors \citep{Hunten+Turco+Toon_1980}.
However, \citet{Hunten+Turco+Toon_1980} estimate
concentrations of only $\sim10^3\cm^{-3}$, or
$\Sigma\sim 10^{-3}\mu{\rm g}\cm^{-2}$, too small to be of
interest here. 

In the lower atmosphere, 
nanoparticles include water droplets and ice crystals,
and products of both natural and anthropogenic combustion.
Droplets will spin like solid spheres, but will presumably lack
the 
large dipole moments required to produce significant rotational absorption.
}

\section{Summary}

Spinning nanoparticles in the ISM with radii $\sim50\Angstrom$
can absorb radiation from the
bright synchrotron background,
producing variations in the brightness
of the radio sky near $80\MHz$ that might be misinterpreted as arising from
21 cm absorption of the CMB at high redshift.

Because the spinning dust absorption would be correlated with other
dust tracers (e.g., FIR emission from the larger grains), such absorption 
could presumably be recognized and corrected for in maps by spatial correlation
with other known astrophysical foregrounds (e.g., emission from dust
at submm and microwave frequencies).


If atmospheric aerosols with radii $\sim70\Angstrom$ are abundant
and have large dipole moments, they could also contribute
absorption below $\sim150\MHz$. Such absorption would of course be
recognizable by its time dependence.

\acknowledgments
This research was supported in part by NSF grant AST-1408723, and in part by
Spanish grants AYA2015-71091-P and MDM-2014-0369.

\bibliography{}
\end{document}